\newcommand{\la}{\lambda}

\newcommand{\be}{\begin{equation}}
\newcommand{\ee}{\end{equation}}

\newcommand{\tht}{\theta}

\documentclass{article}
\usepackage{graphicx}
\usepackage{amsmath}
\usepackage{amsfonts}
\usepackage{amssymb}

\begin{document}
\title{\bf Supersymmetry and Integrability in Planar Mechanical Systems}
\author{Leonardo P. G. de Assis$^{1,2}$ \thanks{E-mail: lpgassis@cbpf.br}$\;$,  Jos\'{e} A. Helay\"{e}l-Neto$^{1,2}$ \thanks{E-mail: helayel@cbpf.br} \\
and Ricardo C. Paschoal$^{3}$ \thanks{E-mail: paschoal@cbpf.br} \\ \\
$^{1}${\it  \normalsize Centro Brasileiro de Pesquisas F\'{\i}sicas -- CBPF,} \\
{\it \normalsize Rua Dr. Xavier Sigaud 150, 22290-180, Rio de Janeiro, RJ, Brasil}  \\
$^{2}${\it \normalsize Grupo de F\'{\i}sica Te\'{o}rica Jos\'e Leite Lopes,} \\
{\it \normalsize P.O.\ Box 91933, 25685-970, Petr\'opolis, RJ, Brasil} \\
$^{3}${\it  \normalsize Servi\c{c}o Nacional de Aprendizagem Industrial, } \\
{\it  \normalsize Centro de Tecnologia da Ind\'{u}stria Qu\'{\i}mica e T\^{e}xtil -- SENAI/CETIQT, } \\
{\it \normalsize Rua Dr. Manoel Cotrim 195, 20961-040, Rio de Janeiro, RJ, Brasil}   }

\date{ }

\maketitle

\begin{abstract}

We present an $N=2$-supersymmetric mechanical system whose bosonic sector, with two
degrees of freedom, stems from the reduction of an $SU(2)$ Yang-Mills theory with the assumption of spatially homogeneous field configurations and a particular ansatz imposed  on the gauge potentials in the dimensional reduction procedure. The Painlev\'{e} test is adopted to discuss integrability and we focus on the r\^{o}le of supersymmetry and parity invariance in two space dimensions for the attainment of integrable or chaotic models. Our conclusion is that the relationships among the parameters imposed by supersymmetry seem to drastically reduce the number of possibilities for integrable interaction potentials of the mechanical system under consideration.
\medskip

           \textbf{PACS}. 11.30.Pb, 11.15.Kc, 05.45.Ac, 02.30.Ik
\end{abstract}


\section{Introduction}

The study of integrability in classical and quantum field theories has been developed
for quite a time, actually since the beginning of the eighties, with relevant results that contributed a great deal for the understanding of these theories and, moreover, allowed the improvement of non-perturbative techniques\cite{int-FT1}--\cite{int-FT2}. On the other hand, a number of streams of investigation on chaos has been pushed forward, mainly considering spatially homogeneous field solutions and by performing calculations in the framework of lattice field theory\cite{chaos-FT1}--\cite{chaos-FT9}. These studies revealed the existence of chaotic solutions in a considerably vast class of gauge theories and, more recently, also in the context of superstrings and supermembrane theories\cite{chaos-STR1}--\cite{chaos-MEMB2}.

Up to now, a detailed analysis relating supersymmetry and chaos, in much the same way as chaos is studied in field theories, is lacking in the literature. Close to this issue, we
should mention a number of attempts to discuss stability and chaos in the framework of brane
theories, by concentrating on their bosonic sector\cite{chaos-MEMB1}--\cite{chaos-MEMB2}.
Nevertheless, even in this context, one should put more emphasis on the specific r\^{o}le of supersymmetry in the determination of stability and chaos. 

A similar situation is observed in connection with the investigation of integrable 
supersymmetric theories, where the integrable or non-integrable character is ascertained,
without however highlighting the mechanisms or those specific properties of supersymmetry which work in favour, or against, integrability\cite{Int-Susy1}--\cite{Int-Susy5}.

Our work sets out to tackle this issue, that we believe should be more manifestly
worked out. To pursue an investigation focusing on the r\^{o}le of supersymmetry in
connection with integrability and chaos, we propose to start off from a supersymmetric
mechanical system, rather than a field-theoretic model.  The system we choose to work with is built up as the $N=2$-extended supersymmetric version of a dimensionally reduced $SU(2)$ Yang-Mills theory that arises when spatially homogeneous fields are considered and a particular ansatz on the gauge potentials is adopted in the dimensional reduction scheme so that only two degrees of freedom survive\cite{Chaos-two} in the mechanical limit. We also devote special attention to the r\^{o}le of parity symmetry, since we assume the latter is an invariance of the interactions involved in the systems we shall be considering. Our analysis of integrability shall therefore rely on our considerations on supersymmetry and parity invariance. They dictate special conditions in the space of parameters so that, instead of having to take by decree special choices of these parameters, as it is usually done, we invoke these two invariances to naturally restrict and select possibilities in parameter space. As a matter of fact, we anticipate that parity may appear in two versions for planar systems, and this point shall be suitably taken care of here.

Our paper is organised as follows. In Section 2, we propose a general 2-dimensional
purely bosonic model with parity symmetry and we identify the cases of integrability.
Next, the $N=2$-supersymmetric extension of the model is written down in Section 3. The complete bosonic sector, now enlarged by the presence of two supersymmetries, is discussed in full details in Section 4, where we pay due attention to the r\^{o}le of parity and we pick out Painlev\'{e} test as a criterium to infer about integrability. In Section 5, we reassess the question of the integrability for the bosonic sector of our $N=2$-model, but now taking into account the constraints dictated by parity whenever it is imposed also to the fermionic interactions. A very restrictive class of potentials comes out that fulfills integrability. Finally, in Section 6, we present our Final Discussions and we draw our General Conclusions.

\section{The ordinary bosonic model with considerations on parity symmetry}

We assume the most general fourth-order polynomial potential for two degrees of freedom
described by the variables $x$ and $y$:
\begin{equation}
V=C_1x^{4}+C_2y^{4}+C_3x^{3}y+C_4xy^{3}+C_5x^{2}y^{2}+C_6x^{3}+C_7y^{3}%
+C_8x^{2}y+C_9xy^{2}+C_{10}x^{2}+C_{11}y^{2}+C_{12}xy .\label{p7}%
\end{equation}

It may be considered as a sort of protopotential used to build up a general
non-supersymmetric polynomial potential up to fourth order. We are bound to
fourth order because we have in mind mechanical models derived from Yang-Mills
theories and these, as we know, display self-interaction vertices for three and four potentials. Since we are interested in realistic models, we impose parity
symmetry which is respected by mechanical and electromagnetic models. We shall not be dealing with models coming from chiral gauge theories.

To implement parity in the model, we have to consider that there are two possibilities, since we are in a 2-dimensional space:
\begin{equation}
x\text{-parity}:%
\genfrac{}{}{0pt}{}{x\rightarrow-x}{y\rightarrow y},%
\end{equation}
or
\begin{equation}
y\text{-parity}:%
\genfrac{}{}{0pt}{}{x\rightarrow x}{y\rightarrow-y}.%
\end{equation}

In the first case, the resulting potential is
\begin{equation}
V=C_1x^{4}+C_2y^{4}+C_5x^{2}y^{2}+C_7y^{3}+C_8x^{2}y+C_{10}x^{2}+C_{11}y^{2} . \label{p10}%
\end{equation}

This potential looks like the sum of two well-known
potentials:

a quartic potential (Yang-Mills-type)
\begin{equation}
V_{YM}=Ax^{2}+By^{2}+ax^{4}+by^{4}+dx^{2}y^{2}, \label{p11}%
\end{equation}
which is known to be integrable in the following cases\cite{Int-Poly}:

\begin{tabular}
[c]{|l|l|}\hline
$a)\text{ \ \ }A=B,\text{ \ \ \ }\ a=b,\text{ \ }d=6a.$ & $\ \ \ that\text{
}in\text{ }our\text{ }case\text{ }is\ \text{\ \ \ }%
C_{10}=C_{11},C_1=C_2,C_5=6C_1.\newline $\\\hline
$b)\text{ \ \ }A,\text{ \ }B,\text{ \ \ \ \ }\ a=b,\text{ \ }d=2a.$ &
$\ \ \ that\text{ }in\text{ }our\text{ }case\text{ }is\ \text{\ \ \ }%
C_{10},C_{11},C_1=C_2,C_5=2C_1.$\\\hline
$c)\text{ \ \ }A=4B,\text{ \ }\ a=16b,\text{ \ }d=12a.$ & $\ \ \ that\text{
}in\text{ }our\text{ }case\text{ }is\ \text{\ \ \ }C_{10}=4C_{11},C_1=16C_2,C_5=12C_1$%
\\\hline
$d)\text{ \ \ }A=4B,\text{ \ }\ a=8b,\text{ \ }\ d=6b.$ & $\ \ \ that\text{
}in\text{ }our\text{ }case\text{ }is\ \text{\ \ \ }C_{10}=C_{11},C_1=C_2,C_5=6C_1.$%
\\\hline
$e)\text{ \ \ }d=0\text{ \ \ }(trivial)$ & $\text{\ \ \ }that\text{ }in\text{
}our\text{ }case\text{ }is\text{\ \ \ }C_5=\text{ }0.$\\\hline
\end{tabular}

 \medskip

and the  Henon-Heiles potential:
\begin{equation}
V_{HH}=\frac{1}{2}\left(  Ax^{2}+By^{2}\right)  +ax^{2}y-\frac{1}{3}by^{3},
\label{p12}%
\end{equation}
that exhibits well-known integrable cases\cite{Int-Poly}:

\begin{tabular}
[c]{|l|l|}\hline
$a)\text{ \ \ }A=B,\text{ \ }\ \ \ \text{\ }a=-b.$ & $\text{ \ \ \ \ }%
that\text{ }in\text{ }our\text{ }case\text{ }is\ \text{\ \ \ }C_{10}=C_{11},\text{
\ }C_7=\frac{1}{3}C_8.$\\\hline
$b)\text{ \ \ \ }A,\text{ \ }B,\text{ \ }\ \text{\ \ \ \ }6a=-b.$ &
$\text{\ \ \ \ \ }that\text{ }in\text{ }our\text{ }case\text{ }%
is\ \text{\ \ \ }C_{10},\text{ }C_{11},\text{ \ \ }C_7=2C_8.$\\\hline
$c)\text{ \ \ \ }16A,\text{ \ }B,\text{ \ }\ 16a=-b.$ & $\text{ \ \ \ \ }%
that\text{ }in\text{ }our\text{ }case\text{ }is\ \text{\ \ \ }C_{10}=16C_{11},\text{
\ }C_7=\frac{16}{3}C_8.$\\\hline
$d)\text{ \ \ \ }a=0\text{ \ }(trivial)$ & $\text{ \ \ \ \ }that\text{
}in\text{ }our\text{ }case\text{ }is\ \text{\ \ \ }C_8\text{ }=\text{ }%
0$\\\hline
\end{tabular}

\section{The supersymmetric model}

Now, we shall consider an $N=2$ supersymmetric mechanical model\cite{SQM}, defined as
follows. The two (complex) Grassmannian parameters of the superspace will be denoted
by $\theta$ and $\overline\theta$. The two real coordinates of a planar particle, $x$ and
$y$, are the bosonic components of the superfields coordinates, which are given by
\be
X(t,\tht,\overline\tht)=x(t)+\Theta^{\dagger}\gamma_{1}\Lambda(t)+\Lambda^{\dagger}(t)\gamma_{1}
\Theta - \frac{1}{2}\Theta^{\dagger}\gamma_{3}\Theta f_{1}(t), \label{spcpx}
\ee
and
\be
Y(t,\tht,\overline\tht)=y(t)+\Theta^{\dagger}\gamma_{2}\Xi(t)+\Xi^{\dagger}(t)\gamma_{2}
\Theta - \frac{1}{2}\Theta^{\dagger}\gamma_{3}\Theta f_{2}(t), \label{spcpy}
\ee
with:
\be \Theta \equiv \binom{\theta}{\overline{\theta}}, \;\;\;\; \Lambda \equiv
\binom{\lambda_{1}}{\lambda_{2}}, \;\;\;\; \Xi \equiv
\binom{\xi_{1}}{\xi_{2}},\label{spcpz}
\ee
where all the $\lambda$'s and $\xi$'s are
Grassmannian variables. The $\gamma_j$'s are the Dirac matrices corresponding to the
two-dimensional Euclidean space under consideration and they may be chosen so as to
coincide with the Pauli matrices: $\gamma_i \equiv \sigma_i$ and $\gamma_3\equiv
-i\gamma_1\gamma_2 = \sigma_3$. $\Theta$ is Majorana spinor, which, in this particular
representation of the $\gamma$-matrices, takes the form given in (\ref{spcpz}), where
the "bar" stands for complex conjugation. On the other hand, $\Lambda$ and $\Xi$ are
Dirac fermions. Therefore, Eqs.~(\ref{spcpx}--\ref{spcpy}) yield:
\be
      X=x+\theta\left(  \lambda_{1}-\overline{\lambda}_{2}\right)  - \overline
{\theta}\left(  \overline{\lambda}_{1}-\lambda_{2}\right)  +\theta
\overline{\theta}f_{1}
\ee
and
\be
      Y=y+i\theta\left(  \xi_{1}-\overline{\xi}_{2}\right)  + i \overline
{\theta}\left(  \overline{\xi}_{1}-\xi_{2}\right)  +\theta
\overline{\theta}f_{2}.
\ee

It is noteworthy to remark that it is precisely the combination $\left(
\lambda_{1}-\overline{\lambda}_{2}\right)  $ the one that carries the
fermionic degrees of freedom of $X$. On the other hand, as for $Y$, its
spinorial degress of freedom are all located in $\left(  \xi_{1}-\overline
{\xi}_{2}\right)  .$

The supersymmetry covariant derivatives are as below:
\begin{eqnarray}
D  &  \equiv & \partial_{\overline{\theta}}-i\theta\partial_{t} \\
\overline{D}  &  \equiv & \partial_{\theta}-i\overline{\theta}\partial_{t},
\end{eqnarray}
they satisfy:
\begin{eqnarray}
D^{2}  &  = & 0 \\
\overline{D}^{2}  &  = & 0 \\
\left\{  D,\overline{D}\right\}   &  = & -2i\partial_{t}.
\end{eqnarray}

The super-action to be considered contains, besides the kinetic terms, the most general superpotential, up to third order in the superfield coordinates (this implies a fourth-order potential in terms of the physical coordinates)
\be
     S = \int{dt d\tht d\overline\tht\: \left\{ \frac{M}{2}\left[ DX \overline{D}X + DY \overline{D}Y \right] + U(X,Y) \right\}},
\ee
where the first term gives rise to the kinetic terms and the superpotential $U(X,Y)$ is assumed to be given by:
\be
       U(X,Y) = k_1 X^2Y + k_2 XY^2 + k_3 X^2 + k_4 Y^2 + k_5 X Y + k'_1X^3 + k'_2Y^3,
\label{spot}
\ee
the $k$'s being arbitrary real constants. Since the term in $XY$ may
be canceled out by means of a proper linear transformation (a rotation in the $X$-$Y$
plane), then the constant $k_5$ may be set as zero, $k_5=0$, without loss of
generality. Similarly, the terms linear in $X$ or in $Y$ were not considered, since
they may be eliminated by a translation redefinition, $X'=X+\mbox{const}$ and
$Y'=Y+\mbox{const}'$. The equations of motion may be used to eliminate the
non-dynamical degrees, of freedom $f_j$, and, thus, the super-action, $S = \int{dt
L}\:$, yields the following Lagrangian where quartic terms in the potential are
present:
\begin{eqnarray}
L & = & \frac{M{\dot{\vec{x}}}^2}{2} +
iM\left( \overline{\la}_j \dot{\la}_j + \overline{\xi}_j \dot{\xi}_j - \overline{\la}_1\dot{\overline{\la}}_2 - \la_2\dot{\la}_1
            - \overline{\xi}_1\dot{\overline{\xi}}_2 - \xi_2\dot{\xi}_1  \right) -
            \frac{k_1^2 +9{k'_1}^2}{2M}x^4 - \frac{k_2^2 +9{k'_2}^2}{2M}y^4 + \nonumber \\*
& & {}-\frac{6k_1k'_1+2k_1k_2}{M}x^3y -\frac{6k_2k'_2+2k_1k_2}{M}xy^3 + \nonumber \\*
& & {}-\frac{2k_1^2+2k_2^2+3k'_1k_2+3k_1k'_2}{M}x^2y^2 -\frac{6k_3k'_1}{M}x^3 -   \frac{6k_4k'_2}{M}y^3 + \nonumber \\*
& & {}-\frac{4k_1k_3+2k_1k_4}{M}x^2y-\frac{4k_2k_4+2k_2k_3}{M}xy^2-\frac{2k_3^2}{M}x^2 -\frac{2k_4^2}{M}y^2 + \nonumber \\*
& & {} + \left[ 2ik_1\left(\la_1\overline\xi_1 - \la_1\xi_2 - \overline\la_2\overline\xi_1 + \overline\la_2\xi_2 +\overline\la_1\xi_1 -\overline\la_1\overline\xi_2 - \la_2\xi_1 + \la_2\overline\xi_2 \right)\right. + \nonumber \\*
& & {} \;\;\;\;\;\; - \left. 2k_2\left( \xi_1\overline\xi_1 - \xi_1\xi_2 - \overline\xi_2\overline\xi_1 +\overline\xi_2\xi_2 \right) -
6k'_1\left( \la_1\overline\la_1 - \la_1\la_2 - \overline\la_2\overline\la_1 +\overline\la_2\la_2 \right) \right] x \nonumber \\*
& & {} + \left[ 2ik_2\left( \overline\la_1\xi_1 - \overline\la_1\overline\xi_2 - \la_2\xi_1 + \la_2\overline\xi_2 +\la_1\overline\xi_1 -\la_1\xi_2 - \overline\la_2\overline\xi_1 + \overline\la_2\xi_2 \right)\right. + \nonumber \\*
& & {} \;\;\;\;\;\; - \left. 2k_1\left( \la_1\overline\la_1 - \la_1\la_2 - \overline\la_2\overline\la_1 +\overline\la_2\la_2 \right) -
6k'_2\left( \xi_1\overline\xi_1 - \xi_1\xi_2 - \overline\xi_2\overline\xi_1 +\overline\xi_2\xi_2 \right) \right] y \nonumber \\*
& & {} - 2k_3 \left( \la_1\overline\la_1 - \la_1\la_2 - \overline\la_2\overline\la_1 +\overline\la_2\la_2 \right)
- 2k_4 \left( \xi_1\overline\xi_1 - \xi_1\xi_2 - \overline\xi_2\overline\xi_1 +\overline\xi_2\xi_2 \right) .
\end{eqnarray}

In the next sections, the integrability conditions for this Lagrangian will be
discussed, and the influence of supersymmetry and parity invariance shall be highlighted.

\section{  The bosonic sector and its integrability.}

The direct application of the the Painlev\'{e} test (for a short review, see Appendix
\ref{alpha}) directly to the bosonic sector is not actually a good procedure, for the
resolution of the systems that appear in the analysis becomes very complex.

In this section, we shall take into consideration the observation
that the original model is not invariant under the two classes of parity
transformations. This may set a more formal framework.

So, in a first attempt, we will impose parity symmetry, that is
a discrete symmetry, only to the bosonic sector of the theory and
after that we shall check how the constraints imposed by this invariance
affects the integrability of the model.

Adopting invariance under $x$-parity, we have the following constraints on the
coefficients of the potential:
\be
C_3=(6  k_1  k'_1+2  k_1  k_2)=0,
\ee%

\be
C_4=(6  k_2  k'_2+2  k_1  k_2)=0,
\ee%

\begin{equation}
C_6=(6  k_3  k'_1)=0, \label{p31}%
\end{equation}%

\be
C_9=(4  k_2  k_4+2  k_2  k_3)=0 ,
\ee%
where the $C$'s above are the coefficients of the bosonic sector of the original potential
for which the parity symmetry is broken.

\subsection{Parameters surviving the parity constraints}

Solving the system of conditions for $k_1,k_2,k'_1,k'_2,k_3$ $\ e$
$\ k_4$,\ we obtain as solution the following possibilities:
\begin{align}
\{k'_1  &  =k'_1,k'_2=k'_2,k_4=k_4,k_3=0,k_2=0,k_1=0\},\label{p34}\\
\{k_1  &  =k_1,k'_2=k'_2,k_4=k_4,k_3=k_3,k'_1=0,k_2=0\},\nonumber\\
\{k'_2  &  =0,k'_1=k'_1,k_2=k_2,k_3=0,k_1=0,k_4=0\},\nonumber\\
\{k'_1  &  =k'_1,k_1=k_1,k_2=-3  k'_1,k'_2=-1/3  k_1,k_3=0,k_4=0\},\nonumber\\
\{k'_2  &  =0,k_2=k_2,k_3=k_3,k'_1=0,k_4=-1/2
k_3,k_1=0\}\nonumber.
\end{align}
 
To study the consequences of these solutions we shall present in
the next subsection the Painlev\'{e}'s test (see Appendix
\ref{alpha}) that have been very used in the search for integrable
systems for being algorithm and with wide application.

\subsection{  Applying the Painlev\'{e} test}

For the first case
$\{k'_1=k'_1,k'_2=k'_2,k_4=k_4,k_3=0,k_2=0,k_1=0\},$ we have the
following potential:
\begin{equation}
Pot_{1}=\frac{9}{2} \frac{{k'_1}^{2}}{M}  x^{4}+\frac{9}{2}
\frac{{k'_2}^{2}}{M}  y^{4}+6  k_4 \frac{k'_2}{M}  y^{3}%
+2 \frac{k_4^{2}}{M}  y^{2} .\label{p35}%
\end{equation}
 
Applying the Painlev\'{e} test, we obtain four branches referring
to the uncoupled systems that survive the test. 
 
For the second case
$\{k_2=0,k'_1=0,k'_2=k'_2,k_4=k_4,k_1=k_1,k_3=k_3\},$ we have the
following potential:
\begin{align}
&  Pot_{2}=\frac{1}{2} \frac{k_1^{2}}{M}  x^{4}+\frac{9}{2}
\frac{{k'_2}^{2}}{M}  y^{4}+\frac{(2  k_1^{2}+3  k_1  k'_2)}{M}
x^{2}  y^{2}+\\
+6  k_4 \frac{k'_2}{M}  y^{3}  &  +\frac{(4  k_1  k_3+2
k_1  k_4)}{M}  x^{2}  y+2 \frac{k_3^{2}}{M}  x^{2}%
+2 \frac{k_4^{2}}{M}  y^{2},%
\end{align}
with dominant potencies:%
\be
\alpha_{1}=-1,\alpha_{2}=-1
\ee
and four branches with the following expressions for the resonances:
\be
-1,\text{ }4,\text{ }\frac{(2  k_1-3  k'_2)}{k_1},\text{ }\frac{(3
k'_2+k_1)}{k_1},%
\ee
that will show integer resonances if we set \ $k_1,$ $k'_2=n \frac{1}%
{3}  k_1$ where $n=\left\{  -1,0,1,2\right\}  $.

For the case $\ n=-1$, it is not possible to determine the resonances.

For the case $\ n=0$, we have the following potential:%
\begin{equation}
Pot_{3}=\frac{1}{2}\frac{k_1^{2}}{M}x^{4}+2\frac{k_1^{2}}{M}x^{2}y^{2}%
+\frac{(4k_1k_3+2k_1k_4)}{M}x^{2}y+2\frac{k_3^{2}}{M}x^{2}+2\frac{k_4^{2}}{M}y^{2}
\label{p36-1}.%
\end{equation}

It does not pass the Painlev\'{e} test because there appears a
compatibility condition that cannot be fulfilled:%
\begin{equation}
-4i\sqrt{2}(18k_1^{2}x_{1}^{2}-5k_4^2-4k_3k_4)=0. \label{p36-2}%
\end{equation}

This equality is indeed satisfied if $k_1,k_3$ $e$ $k_4=0$, \ but
this cancels out the potential.

For the case $\ n=1$, we have the following potential:
\begin{equation}
Pot_{4}=\frac{1}{2}\frac{k_1^{2}}{M}x^{4}+\frac{1}{2}\frac{k_1^{2}}{M}y^{4}%
+3\frac{k_1^{2}}{M}x^{2}y^{2}+2k_4\frac{k_1}{M}y^{3}+\frac{(4k_1k_3+2k_4k_1)}{M}%
x^{2}y+2\frac{k_3^{2}}{M}x^{2}+2\frac{k_4^{2}}{M}y^{2}. \label{p36-3}%
\end{equation}

And now, we obtain four branches with the following resonances:%
\be -1,1,2,4, \ee
but with the following compatibility condition:
\be -2 (-k_4+k_3) M=0, \ee
to be verified in the resonance $j=1$ of the first and of
the second branch. Setting $k_3=k_4$, the system becomes compatible and it passes the
Painlev\'{e} test with only two branches and with the potential
now written
like bellow:%
\begin{equation}
Pot_{5}=\frac{1}{2}\frac{k_1^{2}}{M}x^{4}+\frac{1}{2}\frac{k_1^{2}}{M}y^{4}%
+3\frac{k_1^{2}}{M}x^{2}y^{2}+2\frac{k_3k_1}{M}y^{3}+6k_3\frac{k_1}{M}x^{2}%
y+2\frac{k_3^{2}}{M}x^{2}+2\frac{k_3^{2}}{M}y^{2} \label{p37},%
\end{equation}
with dominant potencies:%
\be
\alpha_{1}=-1,\alpha_{2}=-1.
\ee
 
The values of the resonances for the two branches are:
\be -1,1,2,4, \ee
and for the first branch the coefficients of the dominant terms are:%
\be
x_{_{0}}=\frac{1}{2} \frac{iM}{k_1},y_{_{0}}=\frac{1}{2}  i
\frac{M}{k_1}.%
\ee

For the second branch, the coefficients read as follows:%
\be
x_{_{0}}=-\frac{1}{2} \frac{iM}{k_1},y_{_{0}}=-\frac{1}{2}  i
\frac{M}{k_1}.%
\ee

In the first branch, the arbitrary coefficients are:
\be
y_{_{1}},\text{ }y_{_{2}}\text{ \ and \ \ }y_{_{4}},%
\ee
and the arbitrary coefficients of the second branch are:
\be
y_{_{1}},\text{ \ }y_{_{2}}\text{ \ and\ \ }x_{_{4}},%
\ee
reminding that the variable $t_{0}$ is the fourth arbitrary quantity corresponding to
the resonance $-1$. 

So, the system is of fourth order and possesses four arbitrary
coefficients; therefore, it is integrable.

For the case $\ n=2$, \ we have the following potential:
\begin{equation}
Pot_{6}=\frac{1}{2}\frac{k_1^{2}}{M}x^{4}+2\frac{k_1^{2}}{M}y^{4}+4\frac{k_1^{2}%
}{M}x^{2}y^{2}+4k_4\frac{k_1}{M}y^{3}+\frac{(4k_1k_3+2k_4k_1)}{M}x^{2}%
y+2\frac{k_3^{2}}{M}x^{2}+2\frac{k_4^{2}}{M}y^{2}. \label{p37-1}%
\end{equation}

It was not possible to determine the dominant terms.

For the third case
$\{k_3=0,k_4=0,k_2=k_2,k_1=0,k'_1=k'_1,k'_2=0\},$ we have the
following potential:%
\begin{equation}
Pot_{7}=\frac{9}{2} \frac{{k'_1}^{2}}{M}  x^{4}+\frac{1}{2}
\frac{k_2^{2}}{M}  y^{4}+\frac{(2  k_2^{2}+3  k'_1  k_2)}{M}
x^{2}  y^{2} \label{p38};%
\end{equation}
the expressions for resonances in this case are:
\be
-1,4,\frac{\text{ }(3  k'_1+k_2)}{k_2},\text{ }-\frac{(-2  k_2+3
k'_1)}{k_2},%
\ee
that will show integer resonances if we set $k_2,$ $k'_1=n \frac
{1}{3}  k_2$ where $n=\left\{  -1,0,1,2\right\}  $.

For the case $\ n=-1$, the system passes the test with the
following potential:
\begin{equation}
Pot_{8}=\frac{1}{2}\frac{k_2^{2}}{M}x^{4}+\frac{1}{2}\frac{k_2^{2}}{M}y^{4}%
+\frac{k_2^{2}}{M}x^{2}y^{2} \label{p38-2},%
\end{equation}
with dominant potencies:%
\be
\alpha_{1}=-1,\alpha_{2}=-1
\ee
and the values of the resonances for the two branches:
\be 0,-1,3,4. \ee

For the first branch, the coefficients of the dominant terms are:%
\be
x_{_{0}}=\frac{\sqrt{(-M^{2}-k_2^{2}  y_{0}^{2})}}{k_2},y_{_{0}}=y_{_{0}};%
\ee
for the second branch, they are:%
\be
x_{_{0}}=-\frac{\sqrt{(-M^{2}-k_2^{2}  y_{0}^{2})}}{k_2},y_{_{0}}=y_{_{0}}.%
\ee

In the first branch, the arbitrary coefficients are:
\be
y_{_{0}},x_{3}\text{ \ and \ \ }y_{_{4}},%
\ee
and the arbitrary coefficients of the second branch are:
\be
y_{_{0}},\text{ \ }x_{3}\text{ \ and\ \ }y_{_{4}}.%
\ee

For the case $\ n=0$, the system does not pass the test with the
following potential:
\be
Pot_{9}=\frac{1}{2}\frac{k_2^{2}}{M}y^{4}+2\frac{k_2^{2}}{M}x^{2}y^{2},%
\ee
because there appears the following compatibility condition:
\be 18  i  k_2^{2}  y_{1}^{2}=0. \ee

This equation is satisfied only if $k_2$ equal to zero, but in this case the potential
vanishes, and this is not an interesting situation.

For the case $\ n=1$, the system passes the test with the following potential:%
\begin{equation}
Pot_{10}=\frac{1}{2} \frac{k_2^{2}}{M}  x^{4}+\frac{1}{2} \frac
{k_2^{2}}{M}  y^{4}+3 \frac{k_2^{2}}{M}  x^{2}  y^{2} ,
\label{p39}%
\end{equation}
with the same dominant potencies:
\be
\alpha_{1}=-1,\alpha_{2}=-1,
\ee
and resonances:
\be
-1,1,2,4
\ee
in the two branches.

For the first branch, the coefficients of the dominant terms are:%
\be
x_{_{0}}=-\frac{1}{2}  i \frac{M}{k_2},y_{_{0}}=-\frac{1}{2}
i \frac{M}{k_2};%
\ee
and, for the second branch, the coefficients appear as bellow:%
\be
x_{_{0}}=\frac{1}{2}  i \frac{M}{k_2},y_{_{0}}=\frac{1}{2}
i \frac{M}{k_2}.%
\ee

In the first branch, the arbitrary coefficients are:
\be
x_{_{1}},\text{ }x_{_{2}}\text{ \ and \ \ }x_{_{4}}%
\ee
and the arbitrary coefficients of the second branch are:
\be
y_{_{1}},\text{ \ }y_{_{2}}\text{ \ and\ \ }x_{_{4}}.%
\ee

For the case $\ n=2$, the system does not pass the test with the
following potential:%
\begin{equation}
Pot_{11}=2\frac{k_2^{2}}{M}x^{4}+\frac{1}{2}\frac{k_2^{2}}{M} y^{4}%
+4\frac{k_2^{2}}{M}x^{2}y^{2},%
\end{equation}
because it was not possible to determine the dominant terms.

For the fourth case $\{k'_1=k'_1,k'_2=-1/3  k_1,k_2=-3
k'_1,k_1=k_1,k_3=0,k_4=0\},$ we have the following potential (quartic):%
\begin{equation}
Pot_{12}=5 \frac{k_1^{2}}{M}  x^{4}+5 \frac{k_1^{2}}{M}
y^{4}+10 \frac{k_1^{2}}{M}  x^{2}  y^{2}, \label{p40}%
\end{equation}
with $k'_1=k_1$. The resonances in this case are:
\be
0,-1,3,4
\ee
for the two branches.

For the first branch the coefficients of the dominant terms are:%
\be
x_{_{0}}=\frac{1}{10} \frac{\sqrt{(-10  M^{2}-100  k_1^{2}
y_{0}^{2})}}{k_1},y_{_{0}}=y_{_{0}};%
\ee
and, for the second branch, they are:%
\be
x_{_{0}}=-\frac{1}{10} \frac{\sqrt{(-10  M^{2}-100  k_1^{2}
y_{0}^{2})}}{k_1},y_{_{0}}=y_{_{0}}.%
\ee

In the first branch, the arbitrary coefficients are:
\be
y_{_{0}},\text{ }x_{_{3}}\text{ \ and \ \ }y_{_{4}}%
\ee
and the arbitrary coefficients of the second branch are:
\be
y_{_{0}},\text{ \ }x_{_{3}}\text{ \ and\ \ }y_{_{4}}.%
\ee

For the fifth case $\{k'_2=0,k_2=k_2,k_3=k_3,k'_1=0,k_4=-1/2
k_3,k_1=0\},$ we have the following potential:
\begin{equation}
Pot_{13}=\frac{1}{2}\frac{k_2^{2}}{M}y^{4}+2\frac{k_2^{2}}{M}x^{2}y^{2}%
+2\frac{k_3^{2}}{M}x^{2}+\frac{1}{2}\frac{k_3^{2}}{M}y^{2}. \label{p41}%
\end{equation}

This potential does not pass in the Painlev\'{e} test because the
following compatibility condition appears:%
\be
-3  i  k_3^{2}+18  i  k_2^{2}  y_{1}^{2}=0,
\ee
that is only satisfied if $k_2=k_3=0$, and this eliminates our
potential. Therefore this case does not pass the Painlev\'{e} test.

As this potential is of the quartic type, it is easy to verify
that the result of this Painlev\'{e} analysis is in agreement
with the conditions of integrability for this potential type.

\section{The integrability of the bosonic sector with parity considerations for the complete model.}

As verified in the previous section, by imposing parity to the bosonic sector,
the task of finding integrable cases became less arbitrary, in that the choice 
of the coefficients in the terms of the potential was guided by the argument of parity invariance. In spite of that, it was still necessary to fix by hand the values of
some parameters when applying Painlev\'{e} test to recover the integrable cases we
have listed previously.

In this section, we shall impose the parity symmetry not only to the bosonic sector but
also to the fermionic interactions, and we shall verify to which extent the constraints on
the parameters are able to turn the model integrable without the need of
fixing arbitrarily parameters in the Painlev\'{e} test.

\subsection{Two-component formulation of the fermionic sector}

Since the model is classic and non-relativistic, and defined in a
two-dimensional Euclidean space, $E^{2}$, the covariance group is SO(2). We adopt the
representation below for the Clifford algebra:

Therefore, we adopt:
\begin{equation}
\gamma^{1}=\sigma_{x} ,\label{p42}%
\end{equation}%

\begin{equation}
\gamma^{2}=\sigma_{y} ,\label{p43}%
\end{equation}%

\begin{equation}
\gamma_{3}=-i\gamma^{1}\gamma^{2}=\sigma_{z}, \label{p44}%
\end{equation}
such that:
\begin{equation}
\left\{  \gamma^{i},\gamma^{j}\right\}  =2\delta^{ij}1, \label{p45}%
\end{equation}%

\begin{equation}
\left\{  \gamma^{i},\gamma_{3}\right\}  =0 .\label{p46}%
\end{equation}

For a general spinor,
\begin{equation}
\Psi=\binom{\Psi_{1}}{\Psi_{2}}, \label{p47}%
\end{equation}
the action of SO(2) is as given below:
\begin{equation}
\Psi' = e^{-\frac{i}{2}\omega\sigma_{z}}\label{p50} \Psi ,
\end{equation}
where $\omega$ is the rotation angle; therefore $\Psi^{\dagger}\Psi$ is invariant.

Now, we try to identify $x$-and-$y$-parities in the spinorial space.
 
To do that, we start off from the Dirac equation:%
\begin{equation}
i\gamma^{1}\partial_{x}\Psi+i\gamma^{2}\partial_{y}\Psi=0, \label{p61}%
\end{equation}
to which we impose $x$-parity symmetry:
\begin{align}
\Psi(t;\overset{\rightarrow}{x})\text{ \ }\underrightarrow{\text{
\ \ \ }P\text{ \ \ \ }}\text{ \ }\Psi'(t';\overset{\rightarrow}{x'})  &  =\label{p62}\\
&  =R\Psi(t;\overset{\rightarrow}{x})\nonumber\\
&  =R\Psi(t';-x',y'),\nonumber
\end{align}
where $R$ represents the parity matrix in the spinor space:%
\begin{align}
\gamma^{1}R  &  =-R\gamma^{1}\label{p67}\\
\gamma^{2}R  &  =R\gamma^{2}.\nonumber
\end{align}

Then, our parity matrix may be chosen as%
\begin{equation}
R=\gamma^{2} \label{p68}%
\end{equation}%
and, thus,
\begin{equation}
\Psi'(t';\overset{\rightarrow}{x'})=\gamma^{2}\Psi(t;\overset{\rightarrow}{x}). \label{p69}%
\end{equation}

So, all spinors, up to a phase factor, transform under parity by
means of the $\gamma_{2}$-matrix.

Considering the other possibility, that is, the $y$-parity, one can readily check that
parity is represented by the $\gamma_{1}$-matrix:
\begin{equation}
P\left\{
\begin{array}
[c]{c}%
x\text{ }\rightarrow\text{ }x\\
y\text{ }\rightarrow-y
\end{array}
\right.  \label{p82}%
\end{equation}

\begin{equation}
\left.
\begin{array}
[c]{c}%
\Psi\text{ }\rightarrow\gamma^{1}\Psi\text{ }\\
\\
\Psi'
\left(  t';\overrightarrow{x}\text{ }'\right)  \text{ }\rightarrow\gamma_{1}\Psi\left(  t;\overrightarrow{x}\text{ }\right)
.
\end{array}
\right. . \label{p82x}%
\end{equation}

\subsection{The integrability with the parity constraints from the fermionic Sector}

To include the constraints dictated by $x$- or $y$-parity symmetry for the complete
(bosnic + fermionic) model, we propose to actually carry out the analysis directly in
terms of the superfields (\ref{spcpx}) and (\ref{spcpy}).
Rather than following the lengthy procedure of
considering all the terms of the component-field action, we propose to work without
quitting superspace.

The action of the $x$-parity on the superfields is given by
\be
X\rightarrow-X\text{ \ \ and \ }Y\rightarrow Y\text{,}%
\ee
provided that
\begin{align*}
\Theta\text{ }  & \rightarrow\text{ }\gamma_{2}\Theta,\\
\Lambda\text{ }  & \rightarrow\text{ }\gamma_{2}\Lambda,\\
\Xi\text{ }  & \rightarrow\text{ }-\gamma_{2}\Xi,\\
f_{1}  & \rightarrow\text{ \ \ \ \ \ }f_{1},\\
f_{2}  & \rightarrow\text{ \ }-\text{\ }f_{2}.
\end{align*}

With these parity assignments to the fermions and auxiliary fields, the superfield
coordinates transform under parity exactly as above. Moreover, by virtue of the
specific choice of  $\gamma_{2}$, we have that parity acts on $d\theta$,
$d\overline{\theta}$\ and the covariant derivatives as below:
\begin{align*}
D  & \rightarrow-i\overline{D},\text{ \ \ }\overline{D}\rightarrow iD;\\
d\theta & \rightarrow id\overline{\theta},\text{ \ }d\theta\rightarrow
-id\overline{\theta}.
\end{align*}

With all the prescriptions, the volume element \ $dtd\theta d\overline{\theta }$\
picks\ a minus sign. This means that the kinetic terms are naturally
invariant, but parity symmetry of the potential sets%
\be
k_{1}=k_{3}=k_{4}=k_{2}'=0,
\ee
with $k_{2}$ and $k_{1}'$ non-vanishing.\ \

These parameters constraints are the same as the third set we found when only the
bosonic sector was considered and we found only two integrable cases: Potentials 8 and
10 the we rename now as below:
\begin{equation}
Pot_{susy1-x}=\frac{1}{2}\frac{k_{2}^{2}}{M}x^{4}+\frac{1}{2}\frac{k_{2}^{2}%
}{M}y^{4}+\frac{k_{2}^{2}}{M}x^{2}y^{2},\label{psusy1-x}%
\end{equation}
and
\begin{equation}
Pot_{susy2-x}=\frac{1}{2}\frac{k_{2}^{2}}{M}x^{4}+\frac{1}{2}\frac{k_{2}^{2}%
}{M}y^{4}+3\frac{k_{2}^{2}}{M}x^{2}y^{2}.\label{psusy2-x}%
\end{equation}

So, from all integrable cases found when we considered only bosonic sector, only the
two potentials above preserve $x$-parity under complete model consideration.

On the other hand, if we contemplate $y$-parity symmetry for the whole model, we
have that%
\be
X\rightarrow X\text{ \ \ and \ }Y\rightarrow-Y\text{,}%
\ee
provided that
\begin{align*}
\Theta\text{ }  & \rightarrow\text{ }\gamma_{1}\Theta,\\
\Lambda\text{ }  & \rightarrow-\text{ }\gamma_{1}\Lambda,\\
\Xi\text{ }  & \rightarrow\text{ }\gamma_{1}\Xi,\\
f_{1}  & \rightarrow\text{ }-f_{1},\\
f_{2}  & \rightarrow\text{ \ \ \ \ }f_{2}.
\end{align*}

Also, $D\rightarrow-i\overline{D},$ \ \ $\overline{D}\rightarrow iD,$
$d\theta\rightarrow id\overline{\theta}$ and \ $d\theta\rightarrow
-id\overline{\theta}$.

So, as in previous case, $y$-parity invariance is ensured only for
those superfield monomials that change sign under parity. This then impose:%
\be
k_{2}=k_{3}=k_{4}=k_{1}'=0,
\ee
while  $k_{1}$ and $k_{2}'$ are the only coefficients compatible with $y$-parity invariance.

These constraints on the parameters correspond to only one set of solutions that is
found when only the bosonic setor is considered in connection with the $y$-parity, in a
similar way to what happens for $x$-parity. There are only two integrable cases that we
shall present below:
\begin{equation}
Pot_{susy1-y}=\frac{1}{2}\frac{k_{1}^{2}}{M}x^{4}+\frac{1}{2}\frac{k_{1}^{2}%
}{M}y^{4}+\frac{k_{1}^{2}}{M}x^{2}y^{2},\label{psusy1-y}%
\end{equation}
and
\begin{equation}
Pot_{susy2-y}=\frac{1}{2}\frac{k_{1}^{2}}{M}x^{4}+\frac{1}{2}\frac{k_{1}^{2}%
}{M}y^{4}+3\frac{k_{1}^{2}}{M}x^{2}y^{2}.\label{psusy2-y}%
\end{equation}

So, from all integrable cases found when only the bosonic sector is considered, only
the two potentials above preserve $y$-parity if whole model is analysed.

\section{Final discussions and general conclusions.}

Along the previous sections, we carried out an integrability analysis of the bosonic
sector of the supersymmetric model and we verified the appearance  of integrable cases for both coupled and non-coupled systems.

The coupled cases turn out to be classified into two types: a quartic potential
and a potential that is functionally the superposition of a quartic and a Henon-Heiles potential.

Contrary to the situation where we impose parity symmetry to the
complete action (bosonic and fermionic interactions) and the generated potentials come out totally integrable, without the need of setting integrability constraints, the case in
which parity symmetry is imposed only to the bosonic sector yields
integrable potentials only after we take into account the
constraints that appear in the course of Painlev\'{e} analysis. This means
that, if these constraints are not fulfilled, we will be dealing with
non-integrable potentials and therefore with the possibility of chaos.

For the cases where the potentials have a quartic form, there is no need
to go through a chaos analysis for this has already been discussed in the
literature we have previously referred to.

The cases for which the potentials are given by the superposition of a quartic and a Henon-Heiles form are under consideration and, in a forthcoming work, we shall report
the results of a complete analysis \cite{LRH}.
However, in this section, we shall give an example to illustrate how this type of
non-integrable potential admits order-chaos transition by using the
potential of number 6 of Section(4.2):
\begin{equation}
Pot_{6}=\frac{1}{2}\frac{k_1^{2}}{M}x^{4}+2\frac{k_1^{2}}{M}y^{4}+4\frac{k_1^{2}%
}{M}x^{2}y^{2}+4k_4\frac{k_1}{M}y^{3}+\frac{(4k_1k_3+2k_4k_1)}{M}x^{2}%
y+2\frac{k_3^{2}}{M}x^{2}+2\frac{k_4^{2}}{M}y^{2}. \label{pcuv}%
\end{equation}
 
For this purpose, we make use of Lyapunov characteristic exponent (LCE) and
phase portraits \cite{lya1,lya2} and \cite{pain1}. The Lyapunov exponent is a usefull tool to quantify the
divergence or convergence of initial nearby trajetories for a dynamical
system. In a chaotic system, there is at least one positive Lyapunov exponent,
defined as\bigskip%

\[
\sigma_{i}=\underset{t\rightarrow\infty}{\lim}\ln\frac{d_{i}\left(  t\right)
}{d_{i}\left(  0\right)  }%
\]

where $d_{i}\left(  t\right)  $ is a deformation measure of the small
hypersphere of initial conditions in the phase space along the trajectory.
\ The asymptotic rate of expansion of the largest axis is given by the largest
LCE.   By phase portrait we mean a graph of the dynamical variables in
\ phase space that is used to provide a qualitative insight of the dynamical behavior
of the system under study. The accuracy of our computation was verified by checking if the Hamiltoniam was
conserved during the simulation.

Fixing $k_{1}=10$, $M=k_{3}=k_{4}=1$\ \ the potential acquires the following form:

\bigskip%

\[
V:=50x^{4}+200y^{4}+400x^{2}y^{2}+40y^{3}+60x^{2}y+2x^{2}+2y^{2}%
\]

We calculate de largest $\sigma_{i}$\ and its respectives phase portraits, and
we present two cases for the same set of parameters fixed above, but with
different initial contions. First with  $p_{1}(0)=0.1,p_{2}(0)=0.1,q_{1}%
(0)=0.1,q_{2}(0)=0.0$,  Energy=0.035; it presents regular behavior (see
Figures 1 e 2 below).

\bigskip

\bigskip

\begin{center}
    
  \vspace{0.3cm}
{\par\centering
\resizebox*{0.60\textwidth}{!}{\rotatebox{0}{\includegraphics{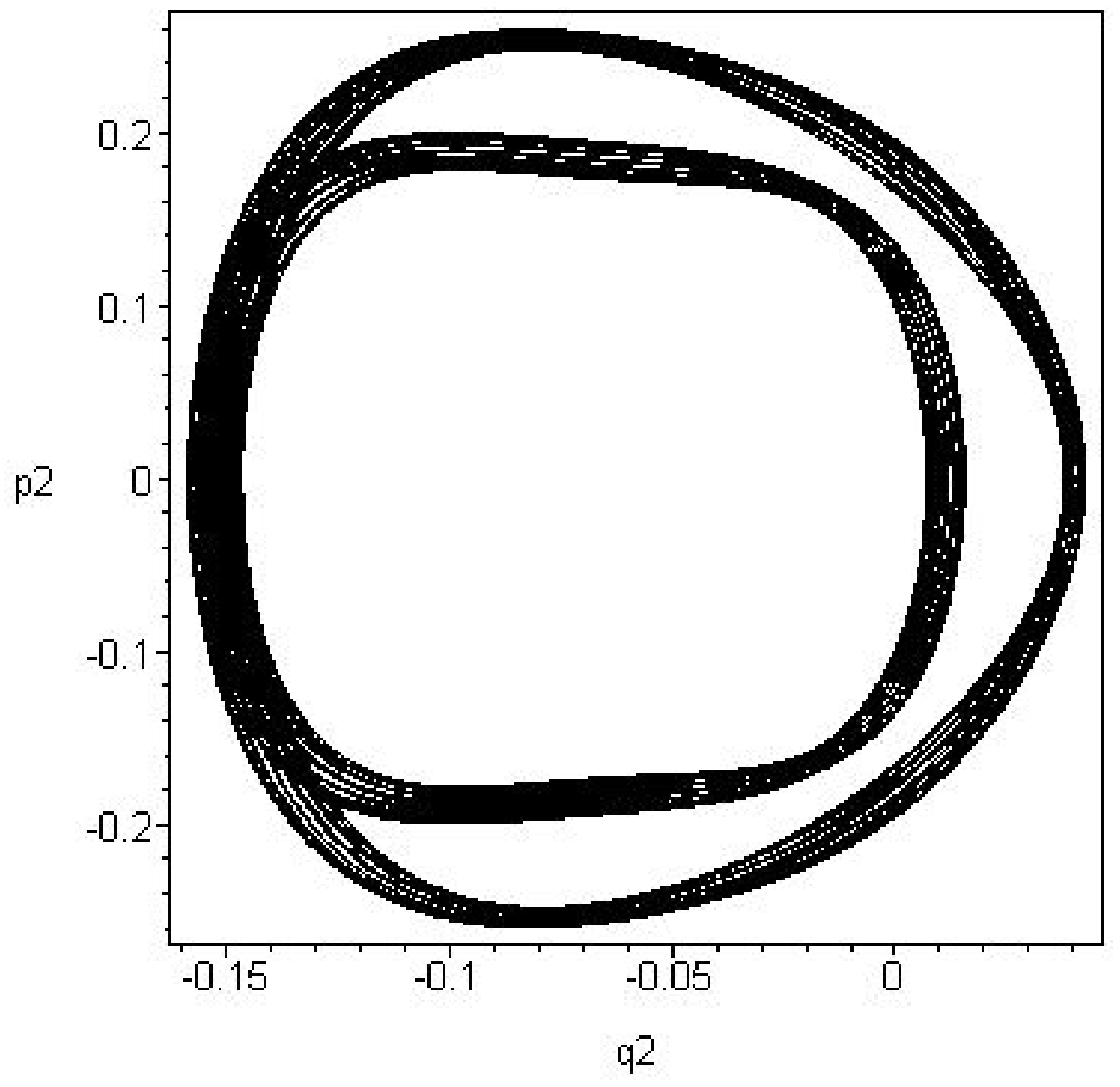}}}
\par}

{\par\centering Figure 1\par}
\vspace{1cm}

\end{center}

\bigskip

\begin{center}
    
  \vspace{0.3cm}
{\par\centering
\resizebox*{0.60\textwidth}{!}{\rotatebox{0}{\includegraphics{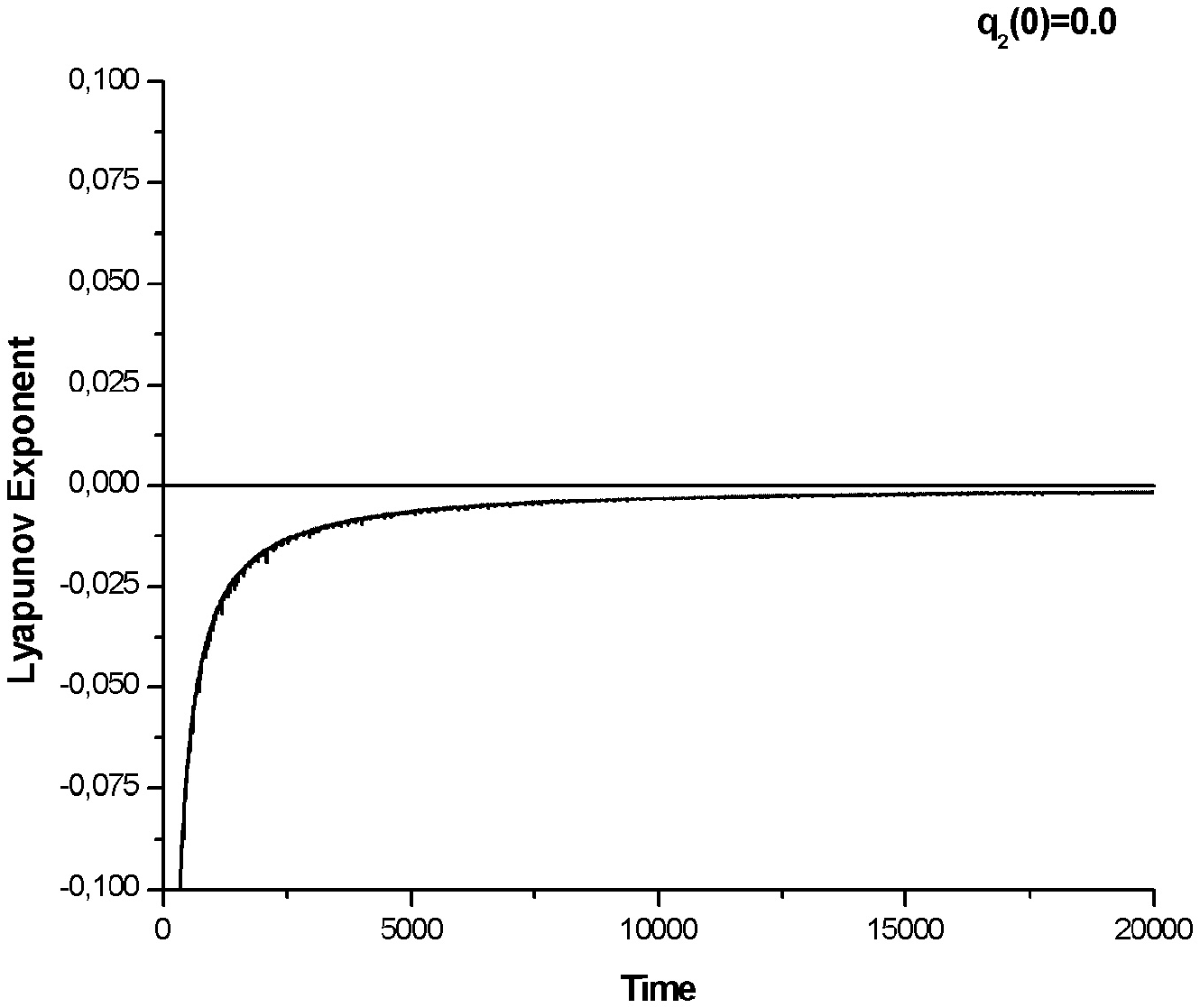}}}
\par}

{\par\centering Figure 2\par}
\vspace{1cm}

\end{center}

\bigskip

The second case is given by \  $p_{1}(0)=0.1,p_{2}(0)=0.1,q_{1}(0)=0.1,q_{2}%
(0)=0.18$, Energy=0.78; it presents chaotic behavior (see Figures 3 e 4 below).

\bigskip

\begin{center}
    
  \vspace{0.3cm}
{\par\centering
\resizebox*{0.60\textwidth}{!}{\rotatebox{0}{\includegraphics{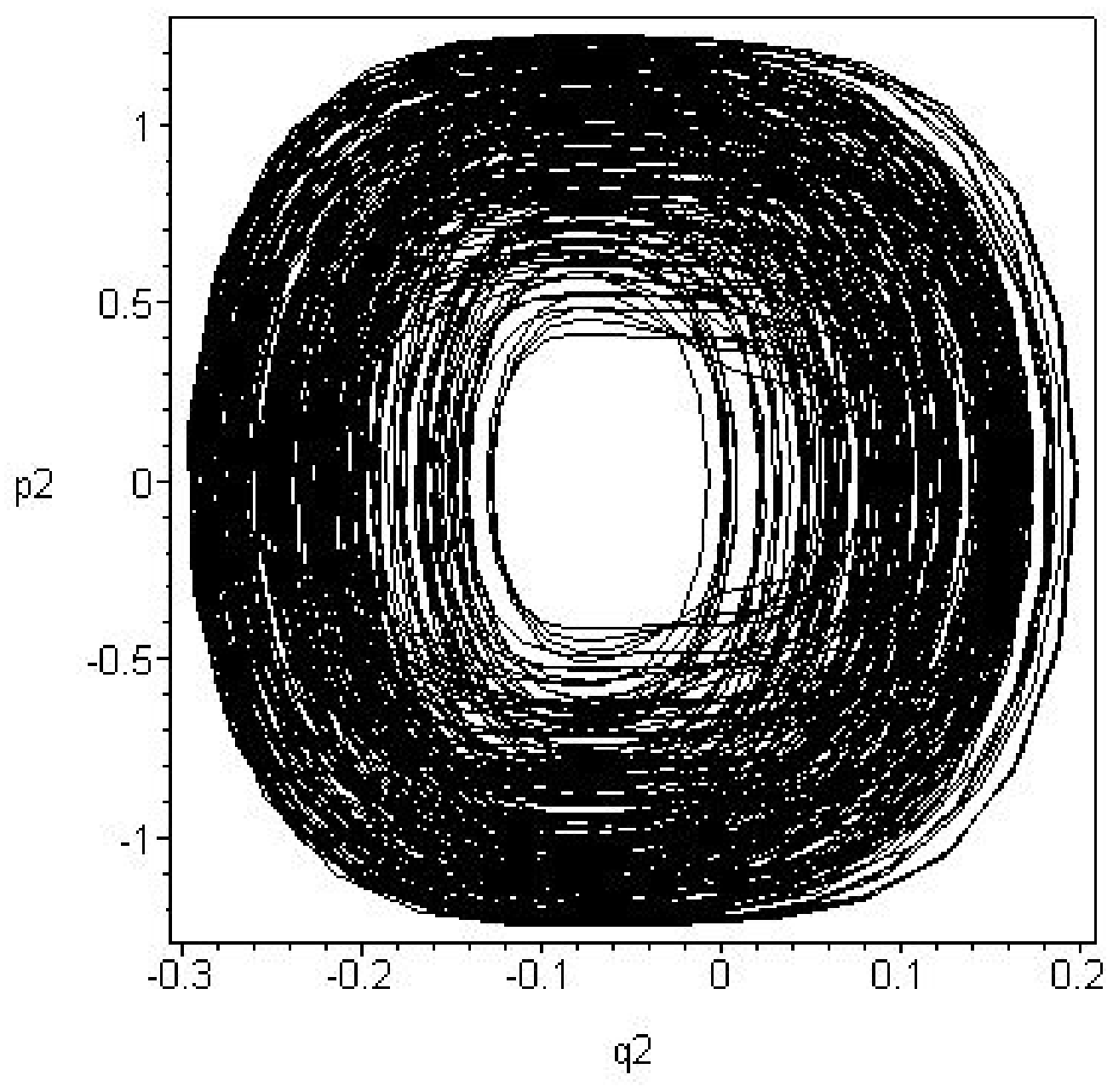}}}
\par}

{\par\centering Figure 3\par}
\vspace{1cm}

\end{center}

\bigskip

\begin{center}
   
  \vspace{0.3cm}
{\par\centering
\resizebox*{0.60\textwidth}{!}{\rotatebox{0}{\includegraphics{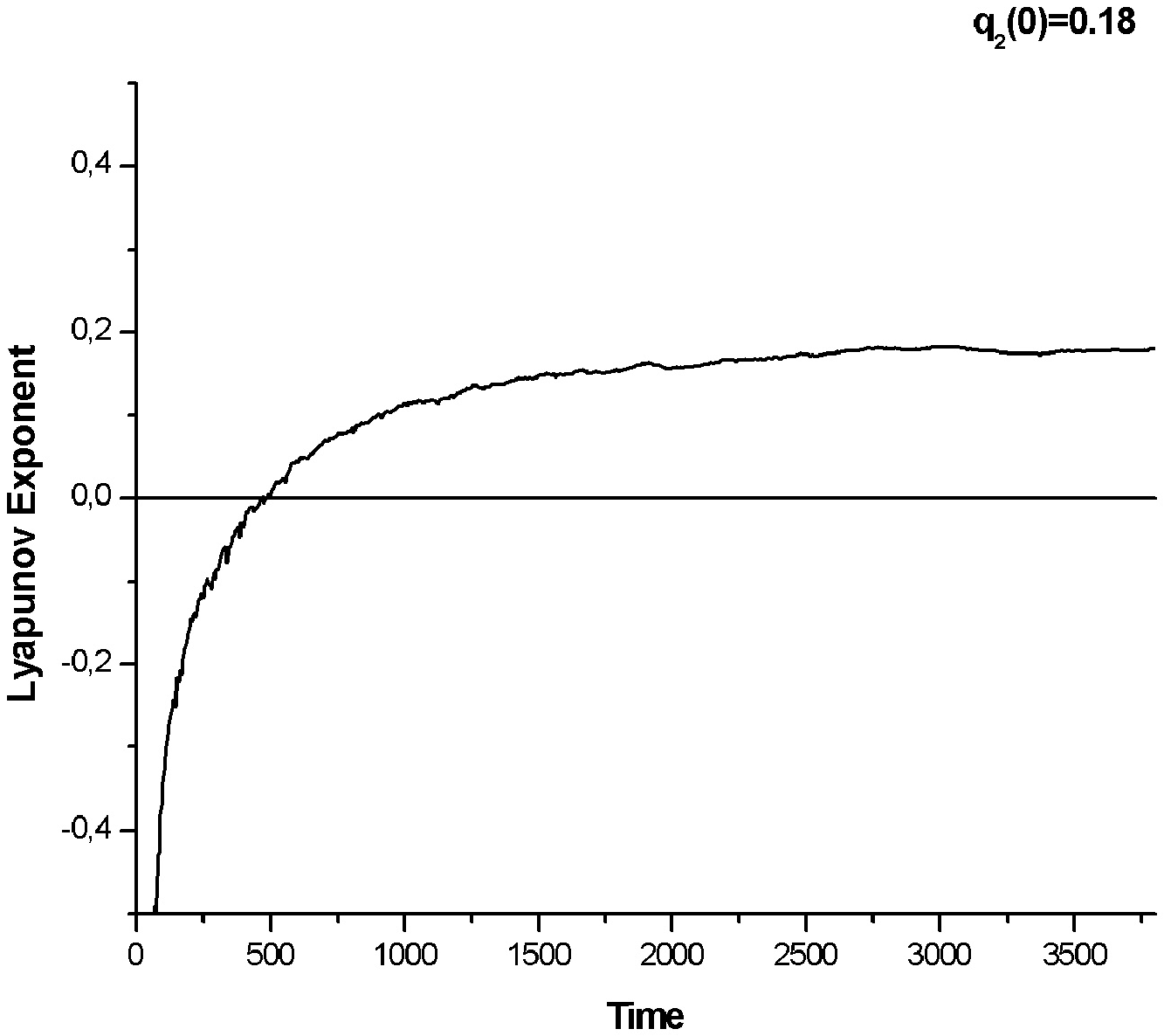}}}
\par}

{\par\centering Figure 4\par}
\vspace{1cm}

\end{center}

\bigskip

It is perhaps noteworthy to say that a potential of the form $\left(
x^{2}+y^{2}\right) ^{2}$ \ \ can be generated by considering the work of
Refs.~\cite{ymmech1}--\cite{ymmech3}, where a classical mechanical Yang-Mills system with
four degrees of freedom $\xi_{i}$ is studied, with Hamiltonian given by
\be
H=\frac{1}{2}\left(  P_{1}^{2}+P_{2}^{2}+P_{3}^{2}+P_{4}^{2}\right)  +\frac
{g^{2}}{8}\left(  \xi_{1}^{2}+\xi_{2}^{2}+\xi_{3}^{2}+\xi_{4}^{2}\right)  .
\ee

Indeed, if we adopt hyperbolic coordinates on the $\left(  \xi_{1};\xi _{2}\right)  $
and $\left(  \xi_{3};\xi_{4}\right)  $ planes,
\begin{align*}
\xi_{1}  & =r_{1}\cosh\theta_{1}  \\
\xi_{2}  & =r_{1}\sinh\theta_{1}  \\
\xi_{3}  & =r_{2}\cosh\theta_{2}  \\
\xi_{4}  & =r_{2}\sinh\theta_{2},
\end{align*}
the Hamiltonian becomes:
\be
H=\frac{1}{2}\left(  P_{r_{1}}^{2}+P_{r_{2}}^{2}-\frac{1}{r_{1}^{2}}%
P_{\theta_{1}}^{2}-\frac{1}{r_{2}^{2}}P_{\theta_{2}}^{2}\right)  +\frac{g^{2}}%
{8}\left(  r_{1}^{2}+r_{2}^{2}\right)^{2}  .
\ee

$\theta_{1}$ and $\theta_{2}$ are ignorable coordinates, so that their
corresponding momenta are integrals of motion. If $P_{\theta_{1}}%
^{2}=P_{\theta_{2}}^{2}=0,$ no negative contributions to the kinetic energy are
present and we get an effective two-degrees-of-freedom system with the potential of
the form  $\left(  x^{2}+y^{2}\right)  ^{2}$ $.$   Besides the discussion of integrability and classical chaos, studies of quantum chaos using this potential have received a great deal of attention in the literature\cite{ymmechQ1}--\cite{ymmechQ6}. This observation may be a good support in favor of the results we have got on the integrable potentials produced after parity symmetry has been imposed for the complete supersymmetric model.
 
\section*{Acknowledgements}
 
The authors thank S.A. Dias for discussions, criticisms and suggestions on an early
manuscript. L.P.G.A. and J.A. H.-N. express their gratitude to CNPq-Brazil for the invaluable financial support.

\appendix
\section{Painlev\'{e} test}
\label{alpha}

The Painlev\'{e} test\cite{pain1}--\cite{pain2} establishes if a system of ODEs exhibits the
Painlev\'{e} property.

An ODE has the Painlev\'{e} property if its solutions in the
complex plane are single-valued in the neighborhood of all its
movable singularities. Given a differential system
\begin{equation}
L_{j}(u_{i},u_{it})=0\text{ \ \ \ \ }\ with\text{ \ \ \ \ }%
i,j=1,...,n\text{\ \ }, \label{p0}%
\end{equation}
we assume a Laurent expansion for the solution
\begin{equation}
u_{i}(t)=(t-t_{0})^{\alpha_{i}}\sum\limits_{k=0}^{\infty}u_{i,k}(t-t_{0})^{k},
\label{p1}%
\end{equation}
with%
\begin{equation}
u_{i,0}\neq0\text{ \ \ \ \ and \ }\alpha_{i}\in Z^{-}\text{\ } ,\label{p2}%
\end{equation}
where $u_{i,k}$ are constants. The algorithm for the Painlev\'{e}
test is implemented by means of the following three steps:

Step 1 (Determine the leading singularity or dominant behavior).
We replace
\begin{equation}
\ u_{i}(t)\simeq u_{i,0}(t-t_{0})^{\alpha_{i}} \label{p3}%
\end{equation}
into (\ref{p0}) to determine $\alpha_{i}$ and \ $u_{i,0}$ and we
obtain an algebraic system with $\alpha_{i}$, assuming negative
integer values and $t_{0}$ arbitrary.

We require that two or more terms of each equation may balance and
determine $\alpha_{i}$ and \ $u_{i,0}.$

If any $\alpha_{i}$\ is not integer, the system is not of
Painlev\'{e} type in its strong version.

If there are more than one solution for $\alpha_{i}$ or $u_{i,0}$\, they define
branches and the following steps of the algorithm need to be applied for each of these
branches.

Step 2 (Determine the resonances).

For each $\alpha_{i}$ and $u_{i,0}$, we calculate the integers $r$
for which $u_{i,r}$ is an arbitrary function in \ref{p0}. We
replace the truncated series
\begin{equation}
\
u_{i}(t)=u_{i,0}(t-t_{0})^{\alpha_{i}}+u_{i,r}(t-t_{0})^{\alpha_{i}+r}.
\label{p4}%
\end{equation}

by ($\ref{p0}$), \ and we look for integer $r$ for which
$u_{i,r}$ is an arbitrary constant .

To do that, after replacing the truncated series by (\ref{p0}), we
keep the most singular terms in $(t-t_{0})$, and the coefficients of
$u_{i,r}$ are set to zero. We get:
\begin{equation}
Q  u_{r}=0,\text{ \ \ }ur=(u_{1,r}\ u_{2,r}...u_{M,r})^{T}, \label{p5}%
\end{equation}
with $Q$ an $M\times M$ matrix depending of $r$.

The resonances are the roots of $\mbox{det}(Q)=0$.

In every system with the Painlev\'{e} property, the resonance ($-1$)
will be present and correspond to arbitrary $(t-t_0)$. \ The
resonance with zero value may also be present, depending of the
number of arbitrary values $u_{i,0}$.

Step 3 (Compatibility conditions and constants of motion).

For every resonance found in the previous step, there is a
compatibility condition which must be verified in order that the
system pass the Painlev\'{e} test. The compatibility conditions are verified by inserting%
\begin{equation}
u_{i}(t)=(t-t_{0})^{\alpha_{i}}\sum\limits_{k=0}^{r_{M}}u_{i,k}(t-t_{0})^{k}
\label{p6}%
\end{equation}
into (\ref{p0}), where $r_{M}$ is the highest positive integer
resonance.

If all these compatibility conditions are satisfied so that they introduce a
sufficient number of arbitrary constants, then the system is said to be of
Painlev\'{e} type.


\begin{thebibliography}{99}
\bibitem{int-FT1}A. Mironov, Phys. Part. Nuclei \textbf{33}, 537 (2002).

\bibitem{int-FT2}A.V. Marshakov, Phys. Part. Nuclei \textbf{30}, 488 (1999).

\bibitem{chaos-FT1}T.S. Biro, N. Hormann, H. Markum, R. Pullirsch, Nucl. Phys.
B-Proc. Suppl. \textbf{86}, 403 (2000).

\bibitem{chaos-FT2}L. Salasnich, J. Math. Phys. \textbf{40}, 4429 (1999).

\bibitem{chaos-FT3}L. Salasnich, Phys. Atom. Nuclei \textbf{61}, 1878 (1998).

\bibitem{chaos-FT4}C. Mukku, M.S. Sriram, J. Segar, B.A. Bambah, S.
Lakshmibala, J. Phys. A-Math. Gen. \textbf{30}, 3003 (1997).

\bibitem{chaos-FT5}M.S. Sriram, C. Mukku, , S. Lakshmibala, B.A. Bambah, Phys.
Rev. D \textbf{49}, 4246 (1994).

\bibitem{chaos-FT6}B. M\"{u}ller, A. Trayanov, Phys. Rev. Lett. \textbf{68}, 3387 (1992).

\bibitem{chaos-FT7}H. Markum, R. Pullirsch, W. Sakuler, Nucl. Phys. B-Proc.
Suppl. \textbf{119}, 757 (2003).

\bibitem{chaos-FT8}B.A. Berg, H. Markum, R. Pullirsch, Phys. Rev. D
\textbf{5909}, art. no.-097504 (1999).

\bibitem{chaos-FT9}S.G. Matinyan, B. M\"{u}ller, Found. Phys. \textbf{27},
1237 (1997).

\bibitem{chaos-STR1}T. Damour, Int. J. Mod. Phys. A \textbf{17}, 2655 (2002).

\bibitem{chaos-STR2}T. Damour, M. Henneaux, Phys. Rev. Lett. \textbf{86},
4749 (2001).

\bibitem{chaos-MEMB1}I.Y. Aref'eva, A.S. Koshelev, P.B. Medvedev, Nucl. Phys.
B \textbf{579}, 411 (2000).

\bibitem{chaos-MEMB2}I.Y. Aref'eva, A.S. Koshelev, P.B. Medvedev Mod. Phys.
Lett. \textbf{A} 13, 2481 (1998).

\bibitem{Int-Susy1}J.M. Evans, J.O. Madsen, Phys. Lett. B \textbf{389},
665 (1996).

\bibitem{Int-Susy2}D. G. Zhang, Phys. Lett. A \textbf{223}, 436 (1996).

\bibitem{Int-Susy3}J.C. Brunelli, A. Das, J. Math. Phys. \textbf{36}, 268 (1995).

\bibitem{Int-Susy4}A. Das, W.J. Huang, S. Roy, Phys. Lett. A \textbf{157}, 113 (1991).

\bibitem{Int-Susy5}L. Hlavaty, Phys. Lett. A \textbf{137}, 173 (1989).

\bibitem{Chaos-two}T.S. Biro,S.G. Matinyan e B.M\"{u}ller, Chaos and Gauge
Field Theory (World Scientific Publishing Co Pte ltd, New Jersey, 1994).

\bibitem{Int-Poly}M. Lakshmanan, R. Sahadevan. Phys. Rep.-Rev. Sec. Phys.
Lett. \textbf{224}, 1 (1993).

\bibitem{SQM}G. Junker, Supersymmetric Methods in Quantum and Statistical
Physics, (Springer, Berlin, 1996).

\bibitem{LRH} L.P.G. de Assis, R.C. Pachoal, J.A. Helay\"{e}l-Neto, work in progress.

\bibitem{lya1}A. Wolf,J.B. Swift, H.L. Swinney, J.A. Vastano, Physica D \textbf{16}, 285  (1985).

\bibitem{lya2}G.Benettin, L. Galgani, A. Giorgilli, J.M. Strelcyn, Meccanica \textbf{15}, 21 (1980).

\bibitem{pain1}M.Tabor, Chaos and Integrability in Non-Linear Dynamics : An
Introduction (John Wiley \& Sons, Inc., New York, 1989).

\bibitem{pain2}M.J. Ablowitz, A. Ramani, H. Segur, Lett. Nuovo Cim.
\textbf{23} (9), 333 (1978).

\bibitem{ymmech1}S. Ichtiaroglou, J. Phys. A-Math. Gen. \textbf{22}, 3461 (1989).

\bibitem{ymmech2}J. Froyland, Phys. Rev. D \textbf{27}, 943 (1983).

\bibitem{ymmech3}J. Karkowski, Acta Phys. Pol. B \textbf{21}, 529 (1990).

\bibitem{ymmechQ1}W.H. Steeb, J.A. Louw, W. Debeer, A. Kotze, Phys. Scr.
\textbf{37}, 328 (1988).

\bibitem{ymmechQ2}B. Eckhardt, G. Hose, E. Pollak, Phys. Rev. A \textbf{39}, 3776 (1989).

\bibitem{ymmechQ3}W.D. Heiss, A.A. Kotze, Phys. Rev. A \textbf{44}, 2403 (1991).

\bibitem{ymmechQ4}M.S. Santhanam, V.B. Sheorey, A. Lakshminarayan, Pramana-J.
Phys. \textbf{48}, 439 (1997).

\bibitem{ymmechQ5}P.L. Christiansen, J.C. Eilbeck, V.Z. Enolskii, N.A. Kostov,
Proc. R. Soc. London Ser. A \textbf{456}, 2263 (2000).

\bibitem{ymmechQ6}M. Tomiya, N. Yoshinaga, Physica E \textbf{18}, 350 (2003).
\end{thebibliography}
\end{document}